\documentclass[aps,twocolumn,showpacs,superscriptaddress]{revtex4}
\usepackage{psfig}
\newcommand{\bea}{\begin{eqnarray}}
\newcommand{\beq}{\begin{equation}}
\newcommand{\eea}{\end{eqnarray}}
\newcommand{\eeq}{\end{equation}}
\begin{document}

\title{Symmetry breaking and the random-phase approximation
in small quantum dots}
\author{Lloren\c{c} Serra}
\affiliation{Departament de F{\'\i}sica, Universitat de les Illes Balears,
E-07122 Palma de Mallorca, Spain}
\author{Rashid G.~Nazmitdinov}
\affiliation{Departament de F{\'\i}sica, 
Universitat de les Illes Balears, E-07122 Palma de Mallorca, Spain}
\affiliation{Bogoliubov Laboratory of Theoretical Physics,
Joint Institute for Nuclear Research, 141980 Dubna, Russia}
\author{Antonio Puente}
\affiliation{Departament de F{\'\i}sica, Universitat de les Illes Balears,
E-07122 Palma de Mallorca, Spain}

\date{April 10, 2003}

\begin{abstract}
The random-phase approximation has been used to compute the properties
of parabolic two-dimensional quantum dots beyond the mean-field 
approximation. Special emphasis is put on the ground state 
correlation energy, the symmetry restoration and the role of the 
spurious modes within the random-phase approximation.
A systematics with the Coulombic interaction strength is presented 
for the 2-electron dot, while for the 6- and 12-electron dots selected 
cases are discussed. The validity of the random-phase approximation  
is assessed by comparison with available exact results.   
\end{abstract}

\pacs{73.21.La, 73.21.-b}
\maketitle


\section{Introduction}

The random-phase approximation (RPA) is a very successful many-body 
theory, widely used to describe Fermi systems in general. It was 
originally proposed by Bohm and Pines  to describe the response
function of the electron gas and, since then, it has been rederived in 
many different ways and contexts \cite{Pines}. 
Important developments of this approximation {\em for finite systems},
including the 
establishment of a formal scheme for both excited and ground states, 
were obtained many years ago in nuclear 
physics \cite{Rowe}.
Unfortunately, great technical 
difficulties have hindered for a long time the solution of the general 
RPA equations, imposing the recourse to severe simplifying assumptions.
In recent times, however, a renewed interest in the RPA as 
a systematic theory of ground state correlations in atomic nuclei has 
appeared (see for instance Ref.\ \onlinecite{Nuclear}).

In the electron-gas theory the RPA has played a benchmark role
in the development of accurate exchange-correlation energy functionals,
specially in the high density limit \cite{Langr}. 
When turning to finite electronic systems,
RPA calculations are generally more involved, due to the lack of 
translational symmetry. Nevertheless, extensive applications to 
metallic clusters (see, e.g., the review by Brack \cite{Brack})
and recent studies of molecules have been performed \cite{Furche}.

A convenient starting point to treat Fermion systems is, in many cases, 
a mean field description like the Hartree-Fock (HF) approach. Selfconsistency
between the mean field and the single-particle orbitals
and total energy minimization are the basic 
conditions at this level. It may happen that the selfconsistent solution
breaks one of the symmetries of the exact Hamiltonian, a well known 
phenomenon called {\em spontaneous symmetry breaking}. It has been 
predicted that the RPA should improve the 
ground state energy. More importantly, it should restore
symmetries which could be broken at the mean field level. 
It is our aim in this work to show with a realistic calculation
for interacting electrons that this is indeed the case. We have chosen 
as an appropriate system a two-dimensional parabolic quantum dot, partly 
because of the interest attracted by semiconductor nanostructures as 
a paradigm of artificial quantum systems, and, also, because these systems 
are believed to be good candidates where electron localization or 
formation of Wigner molecules should be relevant in some regimes
(see Ref.\ \onlinecite{Reimann} for a recent review). We remind that 
Wigner molecules in quantum dots are associated with the breaking
of rotational symmetry of the mean field.

The RPA is, in principle, formulated to address the excited vibrational states,
or {\em vibrons} for short, 
of a Fermion system by means of a superposition of particle-hole 
and hole-particle transitions on 
an a priori unknown correlated ground state. A basic assumption consists in 
treating each particle-hole pair as a boson, i.e., the so-called {\em quasi-boson} 
approximation \cite{Rowe,RS,BR86}. Thus defined, the RPA spectrum describes 
the intrinsic vibrations of the system and, imposing the condition of 
excitation vacuum, also yields an approximation for the correlated ground state. 
In cases of mean field broken symmetry a remarkable property of the RPA is that 
it separates out the collective excitation associated with each broken symmetry 
as a zero-energy or {\em spurious} RPA mode, with an inertial parameter
which is also fixed by the approximation. 
As we shall show below, these spurious RPA modes play a
central role in the restoration of symmetries broken at the mean field
level. The physical excitations of the system 
consist, therefore, of vibrons and collective 
modes associated with spurious solutions, both kinds
being described within RPA. 

It should be noted that the restoration of broken symmetries
can also be attained via projection techniques (see Refs.\ \cite{RS,BR86}).
Examples demonstrating their use for the case of quantum dots have been
recently presented by Yannouleas and Landman \cite{Yan02}.  
In this paper, however, we shall focuss our analysis on the RPA description 
of the ground state and its connection to the excited states. 

The electronic islands formed at the GaAs-AlGaAs interface by using appropriate 
electrodes constitute a remarkable example of controllable quantum system.
Many properties can be described, in a first approach, by using a model 
of electrons confined by an external potential well of parabolic type and 
restricting the electronic motion to the interface plane. These model quantum 
dots have been intensively studied using a variety of methods, including  
semiclassical, mean field and density-functional approaches as well as
exact methods attempting a direct solution of the 
Schr\"odinger equation \cite{Reimann}. 
Within the latter category we find the diagonalization in a basis,
the coordinate-space and the Monte-Carlo solutions. In the exact methods
one normally imposes from the beginning the Hamiltonian symmetries to the solution
and, therefore, the symmetry breaking is not possible (unless degeneracy 
is present). This striking difference with the mean-field solution does not
necessarily imply that the latter one is an artifact. Indeed, the RPA provides
a consistent scheme for the corresponding physical interpretation. 
We attempt in this work a clarification of this issue by comparing in a
test case, such as small quantum dots, exact, mean-field 
and RPA results.

The paper is organized as follows: Section II briefly introduces the
HF mean field approach and discusses the appearance of symmetry-broken 
solutions. Section III is devoted to the RPA excitations, introducing the 
quasi-boson approximation and discussing the RPA spurious mode connected with a 
global rotation when 
the electrons are localized in 2-, 6- and 12-electron dots. 
In Sec.\ IV we discuss the RPA ground state
focussing on the correlation energy and symmetry restoration,
comparing the RPA results with exact values. 
The conclusions are finally drawn in Sec.\ V.
  
\section{The Mean-Field solution}

We consider a system of $N$ electrons whose motion is restricted to 
the $xy$ plane, where a parabolic potential induces the formation of
an electron island. The full Hamiltonian reads
\beq
\label{eq1}
{\cal H} = \sum_{i=1}^{N}
\left[ \frac{{\bf p}^2}{2 m} + \frac{1}{2} m \omega_0^2 r^2 \right]_i
+\sum_{i>j=1}^{N}\frac{e^2}{\kappa\, r_{ij}}\; ,
\eeq  
where $\kappa$ and $m$ are the dielectric constant and electron's effective
mass; $\omega_0$ is the external confinement frequency and we have used 
polar coordinates ($r^2=x^2+y^2$). By introducing the length 
$\ell_0 = (\hbar/m\omega_0)^{1/2}$ and energy  
$E_0 = \hbar^2/m \ell_0^2 = \hbar\omega_0$ units we may rewrite Hamiltonian
(\ref{eq1}) in terms of a single adimensional parameter 
\begin{equation}
R_W= \frac{e^2/(\kappa \ell_0)}{\hbar\omega_0}
\end{equation}
as
\begin{equation}
\label{eq2}
{\cal H} = \sum_{i=1}^{N}{\left[
-\frac{1}{2}\nabla^2 + \frac{1}{2} r^2
\right]_i}
+ R_W \sum_{i>j=1}^{N}\frac{1}{r_{ij}}\; .
\end{equation}  
The $R_W$ parameter was introduced in Ref.\ \onlinecite{rw} as a measure of 
the relative importance of electron-electron interaction to confinement
potential strength. In Ref.\ \onlinecite{YL1} it was used as a measure 
to define the different phases in quantum dots and, in particular, a formation
of Wigner molecules.
In our unit system it is also fulfilled that $\hbar=m=1$.

\subsection{The Hartree-Fock equations} 

Introducing the basis of oscillator and spin states 
that diagonalizes the first term in Eq.\ (\ref{eq2}),
$\{\, |a \eta\rangle;\, a=1,\dots {\cal N};\, \eta=\uparrow,\downarrow\, \}$, 
where $a$ labels the orbital part
and $\eta$ the spin, an arbitrary single-particle 
orbital $|i\rangle$ is expanded as
\begin{equation}
\label{eqir}
|i\rangle = \sum_{a\eta} B^{(i)}_{a\eta}\, |a\eta\rangle\;.
\end{equation}   
The oscillator states $|a\rangle$ 
are characterized by radial ($n_a$)
and angular momentum ($m_a$) quantum numbers --see, e.g., 
Ref.\ \onlinecite{Hawrylak}.
We shall assume that each orbital $i$ has non zero components only for 
a given spin orientation $\eta_i=\uparrow$ or $\eta_i=\downarrow$, i.e.,  
$B^{(i)}_{a\eta}=\delta_{\eta\eta_i}B^{(i)}_{a\eta_i}$. This condition 
imposes good spin symmetry on the single-particle orbitals and it is the 
only symmetry that we shall keep, leaving totally unspecified the 
remaining spatial symmetries. 
Note that the Slater determinant built with these single-particle orbitals
will be an eigenstate of the total $S_z$ operator but not, in general, 
of ${\bf S}^2$ \cite{not2}.
In the chosen basis, the HF equations are written as 
a system of nonlinear eigenvalue equations for the matrix of $B$ coefficients.
The corresponding equation for the $a$-th component of orbital $i$ reads
\begin{eqnarray}
\label{eq5}
\varepsilon^{(0)}_a B^{(i)}_{a\eta_i} &+&
\displaystyle\sum_{c=1}^{\cal N} B^{(i)}_{c\eta_i} 
\left[
\displaystyle\sum_{bd=1}^{\cal N} V_{abcd} 
\left(\displaystyle\sum_{k=1}^{N}{B^{(k)}_{b\eta_k}B^{(k)}_{d\eta_k}}\right)
\right.\nonumber\\
&&\quad - \left.
\displaystyle\sum_{bd=1}^{\cal N} V_{abdc} 
\left(\displaystyle\sum_{k=1}^{N}{\delta_{\eta_i\eta_k}
B^{(k)}_{b\eta_k}B^{(k)}_{d\eta_k}}\right)
\right]\nonumber\\ 
&=& \varepsilon_i B^{(i)}_{a\eta_i}\; .   
\end{eqnarray}  
In these equations $\varepsilon^{(0)}_a$ is the energy of the  
oscillator state $|a\rangle$ and, analogously, $V_{abcd}$ and $V_{abdc}$
are the matrix elements of the 
two-body interaction defined above ($R_W/r_{12}$)
with the corresponding oscillator states.
We can take advantage of the fact that these 
matrix elements can be obtained analytically in terms of the radial ($n_a$) and 
angular momentum ($m_a$) quantum numbers for each basis state $|a\rangle$.

The first and second terms within the square brackets of Eq.\ (\ref{eq5}) 
are the well-known direct and exchange contributions to the HF problem, 
respectively \cite{not1}.
We note that the direct term is independent of spin, while the exchange 
yields a different contribution for $\eta_i=\uparrow$ and $\eta_i=\downarrow$.
Using the selfconsistency approach, both direct and exchange terms are determined
from the 'preceding' iteration and then the eigenvalue problem can be 
separately solved for each spin subset. Successive repetitions of the process
lead eventually to the converged solution. It is worth to stress that this
formulation of the HF equations is particularly well suited to numerical
treatment, since the use of analytical expressions\cite{Chak} for $V_{abcd}$ 
minimizes numerical errors. The only remaining source of numerical
error is, in fact, the precission of the eigenvalue solver. Nevertheless, a 
necessary approximation of the method is the truncation of the basis 
size ${\cal N}$ and one must check the stability of the results with 
this parameter.

\subsection{Some HF results} 

The oscillator basis is defined by including all states below a given
cut-off $E_c$, i.e., $\varepsilon^{(0)}_a<E_c$.
The value of $E_c$ has been chosen large enough to satisfy the above mentioned
convergence criterion. In particular, for $N=2$ and 6 it sufficed to include
the first 55 and 153 oscillator states, respectively. For $N=12$ 
we used up to 190 states which gave good HF convergence. 
With the purpose to 
present in the following sections a systematic comparison of HF, RPA and 
exact results with varying $R_W$ we have taken as test case the $N=2$ quantum dot 
computing its properties for $R_W=0.5$ to 2.5 in steps of 0.5. 
Besides, some selected cases for $N=6$ and 12 have also been calculated.

\begin{figure}[t]
\centerline{\psfig{figure=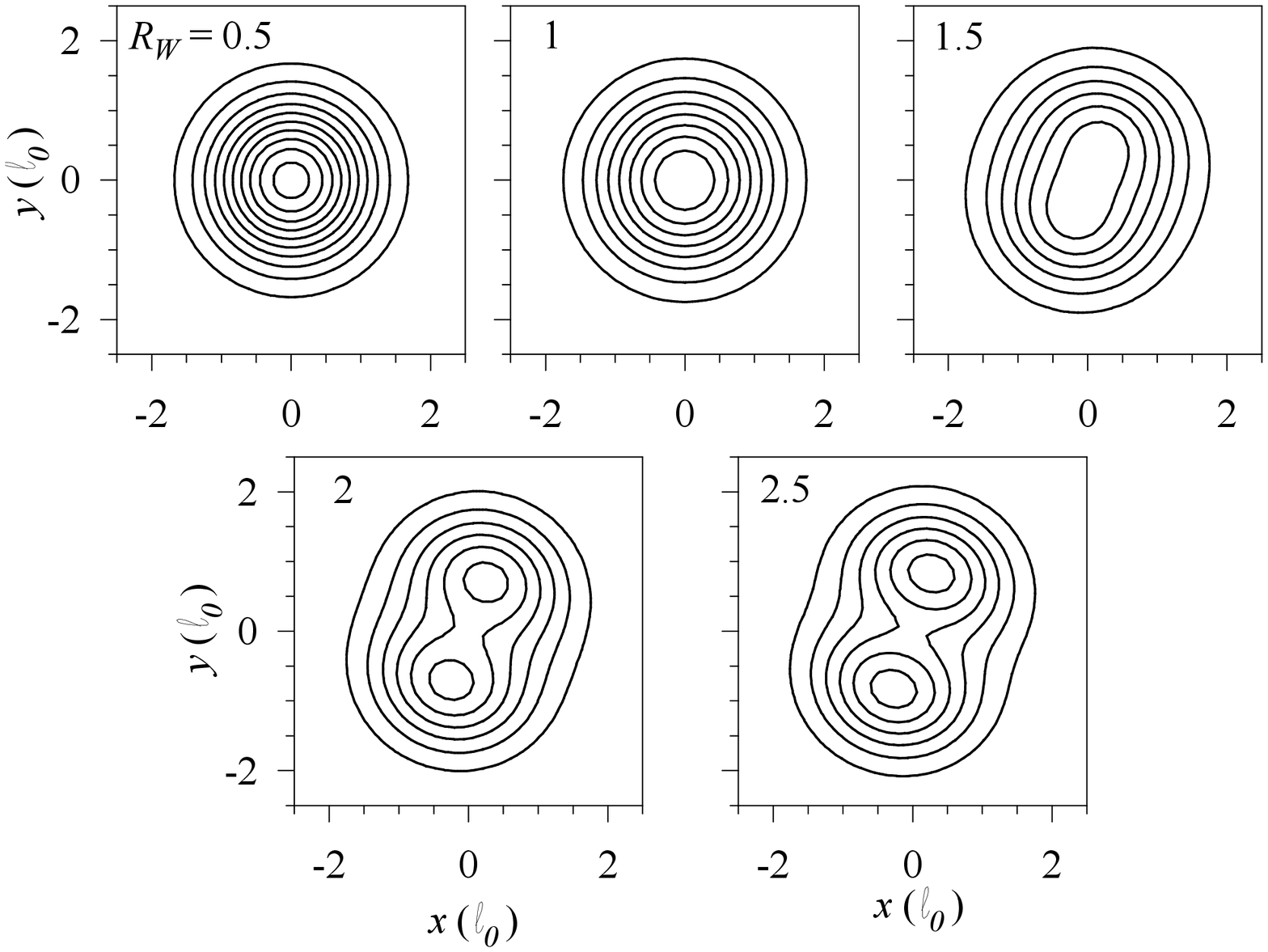,width=3.2in,clip=}}
\caption{
HF densities for the $N=2$ quantum dot with varying $R_W$
parameter (shown in the upper left corner of each panel). From the outermost
contour line inwards each 
line corresponds, respectively, to a density of 0.05, 0.10, 0.15 \dots, 
etc in units of $\ell_0^{-2}$.
The cutoff in the basis has been chosen $E_c\approx 10.6\; E_0$, corresponding
to a basis size ${\cal N}=55$.
}
\label{fig1}
\end{figure}

The HF density 
\beq
\rho_{\rm HF}({\bf r}) = \sum_{i,{\it occ}}{ |\varphi_i({\bf r})|^2 }\; ,
\eeq
where $\varphi_i({\bf r})\equiv\langle{\bf r}|i\rangle$ 
refers to the HF wave functions (\ref{eqir}), are shown 
in Fig.\ \ref{fig1} for the $N=2$ system. We note that between
$R_W=1$ and 1.5 the system changes from a circularly symmetric solution to
a spontaneously broken one, where the two electrons localize in opposite
positions in the mean-field frame. These intrinsic localization becomes
more and more conspicuous as the $R_W$ parameter is increased; an evident
manifestation that if electron repulsion is strong enough 
the favoured solution consists of particles located as far as possible
from each other.

In Fig.\ \ref{fig2} we display the density for the $N=6$ and 12
quantum dots, using $R_W=1.89$. This value has been chosen after Pederiva
{\em et al.} \cite{Pederiva} who, in turn, adjusted it to the 
experiments by Tarucha {\em et al.}\cite{Tarucha}. Both cases show a clear
symmetry breaking; the $N=6$ having the electrons localized on a ring 
while the 12-electron dot has a central electron dimer surrounded by a ring
with 10 electrons. Regarding the HF total energies, we leave the
discussion for the following sections where it will be compared with 
RPA and exact results.

\begin{figure}[t]
\caption{Same as Fig.\ 1 for the $N=6$ and 12 quantum dots. For clarity
the upper plots display a 3d view of the corresponding densities. The 
contour lines are defined as in Fig.\ \ref{fig2} and we have used $R_W=1.89$.
The basis cutoffs for $N=6$ and 12 are $E_c= 16\; E_0$ and $18\; E_0$,
respectively.}
\label{fig2}
\end{figure}

\section{RPA excitations}

This section presents a brief discussion of the main points of 
the RPA for excitations, which are an essential ingredient for the ground 
state description in the same theory. 

\subsection{The Quasi-Boson approximation}

It is assumed that the system's excitations are created by 
operators of the type
\beq
\label{phon}
O_{\lambda}^{\dagger}=\sum_{mi}
{ \left( X_{mi}^{(\lambda)}\, a_m^{\dagger}a_i-
         Y_{mi}^{(\lambda)}\, a_i^{\dagger}a_m \right) }\; ,
\eeq
whose action on the as yet unknown ground state $|0\rangle$
yields the excited vibrational states $|\lambda\rangle$ ($\lambda>0$); i.e.,
\begin{equation}
|\lambda\rangle = O_{\lambda}^{\dagger} |0\rangle\; .
\end{equation}

Hereafter, we shall refer to the electronic vibrational states as 
vibrons. Using the standard notation we refer to
particle states (unoccupied HF orbitals) by subscripts $m$, $n$ and to 
hole states (below the HF Fermi level) by $i$, $j$.
We shall denote the HF Slater determinant for $N$ electrons 
as $|v\rangle$
since it will act as our particle-hole vacuum.
The coefficients $X^{(\lambda)}$ and $Y^{(\lambda)}$ in Eq.\ (\ref{phon}) 
are a set of complex amplitudes characterizing each particular excitation.

The quasi-boson approximation (QBA) amounts to treat each particle-hole
pair ($mi$) as an elementary boson, thus fulfilling
\beq
\label{qba}
[a_i^{\dagger}a_m, a_n^{\dagger}a_j]
\approx
[b_{mi},b_{nj}^\dagger] = \delta_{mn} \delta_{ij}\;,
\eeq
where the first equality introduces the boson operators 
$b^\dagger_{mi}=a^\dagger_ma_i$. 
The result (\ref{qba}) is exact in the average, i.e., when considered
as expectation value with the uncorrelated vacuum $|v\rangle$ as
$\langle v|[a_i^{\dagger}a_m, a_n^{\dagger}a_j]|v\rangle=\delta_{mn}\delta_{ij}$;
see Refs.\ \onlinecite{Rowe,RS,BR86}. 
In this approximation, the creation $O^\dagger_{\lambda}$ and 
annihilation $O_{\lambda}$ operators obey bosonic commutation
relations
\beq
[O_{\lambda},O_{\lambda^\prime}^\dagger] = \delta_{\lambda\lambda^\prime}
\; .
\eeq

At the RPA level the eigenstates and eigenvalues are obtained 
by simplifying the exact equations of motion 
\beq
  \label{eqm}
  \left[ {H}, {O}^\dagger_\lambda \right] = \hbar\omega_\lambda
  {O}^\dagger_\lambda
\eeq
for the vibron operators ${O}^\dagger_\lambda$ and 
excitation energies $\hbar\omega_\lambda$ within the particle-hole 
space.
With the aid of the QBA the particle-hole part of the commutator
$\left[ {H}, {O}^\dagger_\lambda \right]$ can be derived \cite{Rowe,RS,BR86}
and Eq.\ (\ref{eqm}) then transforms to a generalized eigenvalue 
problem, well known in the literature. We stress that henceforth
all commutators are to be understood in the above QBA.
The RPA equations read
\beq
  \label{rpa}
  \left(
    \begin{array}{ll}
      A & B\\
      B^* & A^*
  \end{array}
  \right)
  \left(
  \begin{array}{r}
  X\\
  Y
 \end{array}
  \right)=\hbar\omega_\lambda
  \left(
  \begin{array}{r}
  X\\
  -Y
 \end{array}
  \right)\; ,
 \eeq
 where the submatrices $A$ and $B$ are given by
 \bea
\label{mA}
 A_{mi\,nj}&=&
\varepsilon_{mi}\,\delta_{mn}\delta_{ij} + V_{mjin} - V_{mjni} \nonumber\\
\label{mB}
 B_{mi\,nj}&=& V_{mnij} - V_{mnji}\; .
 \eea
Here, $\varepsilon_{mi}$ is the energy of each bosonic
pair, given in terms of the HF eigenvalues as  
$\varepsilon_{mi}=\varepsilon_m-\varepsilon_i$.

The matrix elements 
in Eqs.\ (\ref{mA}) may be computed in the 
oscillator basis using expressions similar to those of the 
HF eigenvalue problem. For instance, 
\begin{eqnarray}
V_{mjin} &=& \delta_{\eta_m\eta_i}\delta_{\eta_n\eta_j}\,\times\nonumber\\
&& \sum_{abcd=1}^{\cal N}{
V_{abcd}\,
B^{(m)*}_{a\eta_m}
B^{(j)*}_{b\eta_j}
B^{(i)}_{c\eta_i}
B^{(n)}_{d\eta_n}}\; .
\end{eqnarray}

In practice, the numerical solution of Eq.\ (\ref{rpa}) requires the 
diagonalization of a rather large matrix, whose dimension depends on the 
number of hole $N_h$ and particle $N_p$ states. To be consistent with 
our imposed spin symmetry we restrict to bosonic pairs where each member
of the pair has the same spin. 

\subsection{Spurious modes}

It was proved by Thouless \cite{Th} that, when the HF solution corresponds 
to a minimum in the energy surface, the RPA Eq.\ (\ref{rpa}) 
provides only real frequencies $\omega_{\lambda}$. 
Furthemore, if the full Hamiltonian has any symmetry which is broken 
by the mean field there is a corresponding RPA mode with 
zero frequency (a spurious mode) which is orthogonal  
to the other vibrational (nonzero energy) modes.
In other words, the generators of symmetries broken on the mean 
field level create eigenstates with zero energy in RPA.

As shown below, we shall be concerned with rotational symmetry
for which the infinitesimal generator is the orbital angular 
momentum $L_z$.  
To treat the spurious mode related to rotation 
it is convenient to introduce the canonical conjugate operators
$L_z$ and 
an {\em angle} operator $\Phi$; the latter defined by the 
following relations \cite{Th,MW69} 
\bea
\left[ H , {{L}_z} \right] &=&0, \nonumber\\
\label{TV2}
[H,{C}]\; &=& \hbar {L}_z \nonumber\\
\label{TV5}
[{L}_z,{C}] &=& \hbar {\cal J}\; .
\eea
where $C=i{\cal J}\Phi$ is an anti-Hermitian operator 
and ${\cal J}$ is the moment of inertia.

When the above Eqs.\ are
implemented in the RPA they lead to the so-called Thouless-Valatin
moment of inertia\cite{TV62} ${\cal J}_0$ and to the two RPA vectors 
\bea
L_{z} &\equiv& \sum_{mi}
{ \left( \ell_{mi}^{(z)}\, b^{\dagger}_{mi} +
         \ell_{mi}^{(z)*}\, b_{mi} \right) }\; ,\nonumber\\
\label{spur2}
C &\equiv& \sum_{mi}
{ \left( c_{mi}\, b^{\dagger}_{mi} -
         c_{mi}^{*}\, b_{mi} \right) }\; .
\eea
The coefficients of the RPA $L_z$ operator are directly given by
the single-particle HF matrix elements. In contrast,
for the $C$ operator one needs to solve the linear system of equations 
\beq
  \label{rpasp}
  \left(
    \begin{array}{ll}
      A & B\\
      B^* & A^*
  \end{array}
  \right)
  \left(
  \begin{array}{c}
  c\\
  c^*
 \end{array}
  \right)=\hbar
  \left(
  \begin{array}{c}
  \ell^{(z)}\\
  \ell^{(z)*}
 \end{array}
  \right)\; .
 \eeq
Once these two sets of coefficients are determined,
the Thouless-Valatin moment of inertia ${\cal J}_0$
may be calculated as 
\beq
{\cal J}_0=\sum_{mi}{
\left( \ell_{mi}^{(z)*} c_{mi} + \ell_{mi}^{(z)} c_{mi}^* \right)
} \; .
\eeq

The operators of Eqs.\ (\ref{spur2}), together with
the RPA vibrons $O^\dagger_\lambda$, $O_\lambda$, form a complete set 
for any operator linear in the bosons $b_{mi}^{\dagger}$ and $b_{mi}$.
From the point of view of its Fermionic character this could represent
an arbitrary one-particle-one-hole operator of the type
\beq
F = \sum_{mi}{f_{mi} b^\dagger_{mi}+ f_{im} b_{mi}}\; .
\eeq
The corresponding expansion in terms of RPA excitations (for a single spurious mode)  
reads
\bea
\label{def}
{F} & = & 
\sum_{\omega_\lambda>0}\left( [O_{\lambda},{F}]\, O_{\lambda}^{\dagger} + 
                     [{F},O_{\lambda}^{\dagger}]\, O_{\lambda}
\right) \nonumber\\
&+& \frac{1}{{\cal J}_0}\left( [{F},{C}]\, {L}_z + 
                              [{L}_z,{F}]\, {C} \right)\; .
\eea
When applied to the boson operators themselves, the above result
implies that we can expand each $b^\dagger_{mi}$ and $b_{mi}$ in terms of the 
vibrons ($O^\dagger_\lambda$, $O_\lambda$) and spurious modes
($L_z$,$C$).
We shall return to this point in more detail in the next section. 

\subsection{Strength function and sum rules} 

The strength function for a general Hermitian single-particle operator 
$F$ reads
\beq
\label{sf}
S(E)=\sum_{\lambda}\delta(E-\hbar\omega_{\lambda})\,
\left\vert\,\langle\lambda|\,F\,|0\rangle\, \right\vert^2\; .
\eeq
Important quantities related to the strength function are its
energy moments, usually known as sum-rules. The $k$-th moment 
$S_k$ is given by 
\beq
\label{sr}
S_k=\sum_{\lambda}
(\hbar\omega_{\lambda})^k\,
\left\vert\,\langle\lambda|\,F\,|0\rangle\, \right\vert^2\; .
\eeq

It is well known that the RPA fulfills several exact sum 
rules (see Ref.\ \onlinecite{Lippa} for recent applications 
of sum rules to quantum dots).
In particular, for the dipole operators $F=\sum_i{x_i}$ or 
$F=\sum_i{y_i}$ one has 
$S_1 = \hbar^2 N/(2m)$. More importantly, 
the RPA also fulfills the Kohn theorem for parabolic 
confinement (see, e.g., Ref.\ \onlinecite{Chak}) stating 
that the only allowed dipole excitation is the 
center of mass rigid motion at an energy $\hbar\omega_0$. 
These exact results constitute good checks of selfconsistency between 
HF and RPA, as well as of convergence with space size, in numerical 
calculations. 

\subsection{Calculated spectra}

\begin{figure}[t]
\centerline{\psfig{figure=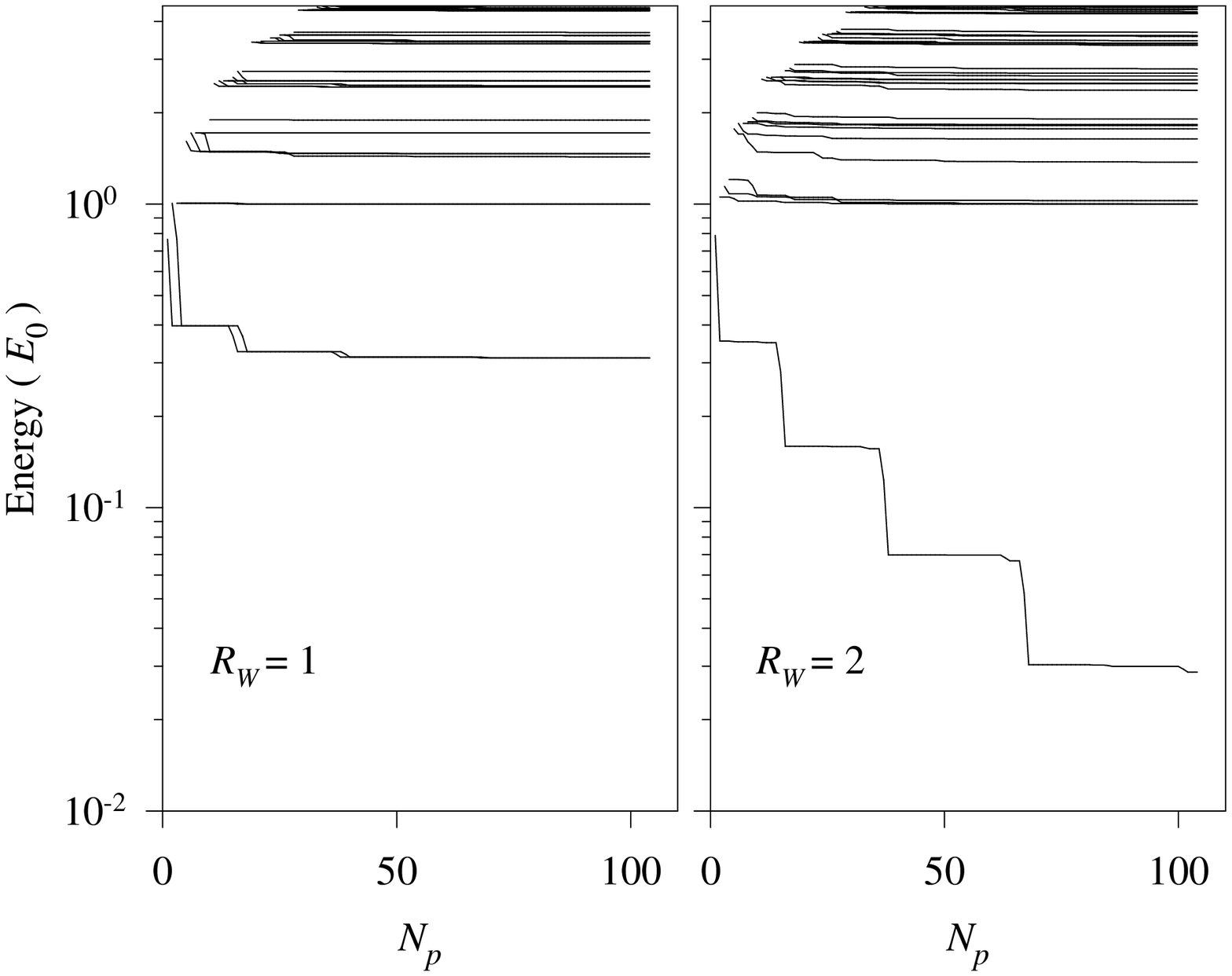,width=3.2in,clip=}}
\caption{Evolution of the RPA excitation spectra with the number of particle 
states included. The results correspond to $N=2$ with $R_W=1$ (left)
and 2 (right). The same scale has been used in both panels for a better 
comparison. Note that increasingly higher RPA excitations appear
as $N_p$ is increased.
The same basis size and cutoff of Fig.\ \ref{fig1} have been used.}
\label{fig3}
\end{figure}

Figure \ref{fig3} displays the eigenvalues $\hbar\omega_\lambda$ for two cases
of the $N=2$ quantum dot as a function of the number of particle states
included in the RPA equations.
The first case ($R_W=1$) is circularly symmetric
at the mean field level while the second one ($R_W=2$) has rotational
broken symmetry (cf.\ Fig.\ \ref{fig1}). The evolution of the eigenvalue set with 
$N_p$ shows a remarkable difference in these two cases.
While all the eigenvalues of the circular dot stabilize for 
a high enough value of $N_p$, the symmetry broken solution exhibits 
one state whose
energy keeps decreasing as the number of particle states is increased. 
It clearly corresponds to the appearance of an RPA spurious mode
connected with the broken rotational symmetry,
which in the limit $N_p\to\infty$ should lie at zero energy. 
Though stable, the first excited state of the
circular case lies rather low in energy, a hint of the proximity
in $R_W$ value to the broken symmetry transition point. Actually, 
for $R_W=0.5$ the  first vibron settles down at a higher energy,  
$E\approx 0.7 E_0$.

Figure \ref{fig4} shows the corresponding spectra for
the $N=6$ and 12 quantum dots. Both HF solutions 
correspond to broken symmetry cases (cf.\ Fig.\ \ref{fig2})
and, therefore, their RPA spectra should display spurious solutions. 
This is clearly the situation for $N=6$, with a well separated low
energy mode. For $N=12$, however, the situation is less convincing and
we actually obtain two modes which seem to separate from the higher
vibron states. We stress that for the 12 electron dot the strong increase
in number of particle-hole pairs prevents us from extending the calculation
to higher $N_p$ values. Nevertheless, a tentative interpretation
may be given taking into account the mean field structure 
of this dot shown in Fig.\ \ref{fig2}. 
One would expect a real spurious mode, corresponding to the 
rigid rotation of the full system, and a soft (low energy) mode 
corresponding to the rotation of the inner electron dimer with
respect to the outer ring of electrons. These two modes could thus 
explain the behaviour of the two lower RPA eigenvalues in the right 
panel of Fig.\ \ref{fig4}.

\begin{table}[b]
\begin{tabular}{cc|c}
$N$ & $R_W$ & ${\cal J}_0$ ($m\ell_0^2$)\\
\hline\hline
2 & 1.5 & 0.68\\
2 & 2 & 1.26 \\
2 & 2.5 & 1.63 \\
6 & 1.89 & 16.4 \\
12 & 1.89 & 57.6\\
\hline
\end{tabular}
\caption{Moments of inertia for the spurious rotational modes obtained
in the Thouless-Valatin approximation (see Sec.\ III-B).}
\end{table}

We analyze next the strength function results. Figure \ref{fig5} displays some 
representative cases, for the dot with six electrons. The upper panels
show the electric ($x$) and magnetic dipole ($\ell_z$) strength 
functions computed within the HF method, while the lower ones are 
the analogous results within the RPA. In HF all excitations are
above the gap between the Fermi level and the first unoccupied
orbital (the HOMO-LUMO gap) which for $N=6$ takes the value
$\approx 1.8 E_0$. In RPA, however, one obtains
states below this gap. Actually, the strength function for the 
$\ell_z$ operator concentrates almost exclusively on the first RPA 
eigenvalue, at a very small energy; a clear manifestation
of rotational spurious mode discussed above. For the dipole
operator there is again a single mode, this time at an energy
$\approx 1 E_0$, and the total sum rule $S1$ is $\approx 3 E_0\ell_0^2$. 
As also discussed above, this proves that our results fulfill Kohn 
theorem with an excellent accuracy.

The Thouless-Valatin moments of inertia obtained for the quantum dots
with spurious modes are summarized in Tab.\ I. These inertial parameters 
define physical rotational bands, characterized by the energies
${\hbar^2 M^2}/(2{\cal J}_0)$ with $M=0, 1, \dots$ \cite{not3}.
We have also checked numerically that the Thoules-Valatin moment of inertia 
coincides, to a good accuracy, with the value obtained from a constrained 
HF calculation for 
${\cal R}={\cal H}-\lambda L_z$ as
\beq
{\cal J}_0 = -\frac{d^2 \langle {\cal R}\rangle }{d\lambda^2}
= \frac{d\langle L_z\rangle}{d\lambda}
\; .
\eeq 
We stress that the equivalence between the two moments of inertia
can be fulfilled {\em if and only if} a self-consistent HF minimum
solution is found.

\begin{figure}[t]
\caption{Same as Fig.\ \ref{fig3} for the $N=6$ and 12 quantum dots.}
\label{fig4}
\end{figure}

\begin{figure}[t]
\caption{HF and RPA excitation cross sections for the $N=6$ quantum dot.
Vertical bars show in an arbitrary scale the position and height 
for each state while the solid line displays the accumulated contributions
to the $S1$ sum rule. Left and right panels correspond to the dipole ($x$)
and rotation ($\ell_z$) operators, respectively.
The inset in the bottom-right panel shows an enlarged view, using logarithmic
scales, of the RPA $\ell_z$ cross section.}
\label{fig5}
\end{figure}

\section{The ground state in the RPA}

The RPA ground state $|0\rangle$ is defined from the condition that
it is the vacuum for all vibrons
\beq
\label{cond1}
O_{\lambda}|0\rangle=0\; .
\eeq
In this section we shall discuss the following aspects of the 
RPA ground state: i) total energy;  
ii) restoration of the HF broken symmetries.

\subsection{Correlation energy}

The total energy in the RPA ($E_{\rm RPA}$) 
can be split in a mean field contribution ($E_{\rm HF}$) and 
a correction ($\Delta_{\rm RPA}$), as
\beq
\label{etot}
E_{\rm RPA} = E_{\rm HF} - \Delta_{\rm RPA}\; .
\eeq
Note that the correction $\Delta_{\rm RPA}$, with a minus sign,
gives the standard correlation energy of the system. A decrease
in ground state energy ($\Delta_{\rm RPA}>0$) will imply an improvement
with respect to the mean field theory. However, since RPA 
is not based on energy minimization, it is not bound to fulfill 
the variational principle and, therefore, $E_{\rm RPA}$ could be even lower
than the exact energy. In other words, the correlation energy could 
be overestimated in RPA, as it has been suggested in the literature
\cite{BR86}.   

The RPA correction reads \cite{RS,BR86}
\beq
\label{erpa}
\Delta_{\rm RPA} = \frac{1}{2}\left(\, 
{\rm Tr}A
-
\sum_{\omega_\lambda>0} \hbar\omega_{\lambda} 
 \, \right) \; .
\eeq
The above Eq.\ (\ref{erpa}) includes the contribution from the vibrons
at positive frequencies $\omega_\lambda$ and, also, from the spurious 
mode associated with a rotational excitation.
Note that Eq.\ (\ref{erpa})
is the result of a partial cancellation between two large terms. In practice, 
this may cause $\Delta_{\rm RPA}$ to converge rather
slowly with space dimension, i.e.,  
a high number $N_p$ of HF particle orbitals may be needed (see below).

\subsection{Numerical results for ground-state energies}

We have analyzed the RPA ground state for the same dots whose excited states
have been discussed in the preceding sections. Figure \ref{fig6} displays 
the evolution of the ground state energy $E_{\rm RPA}$ with number of particle 
states $N_p$. 
The convergence with $N_p$ depends on each specific case: while the $N=2$,
$R_W=1$ energy converges fast enough, the $N=12$ result is clearly still increasing
for the maximum $N_p$ used in the numerics. This is not surprising, since 
one naturally expects that the larger $N_h$ the larger the space requirements
for RPA convergence. A similar conclusion may be drawn regarding the 
intensity of the interaction, as given by the parameter $R_W$. 
In view of the obtained results, $E_{\rm RPA}$ must be 
extrapolated, in general, from the range of the computationally feasible 
values. We have empirically found that a 4-parameter function, 
$a- b/(1+cN_p)^{1/d}$, leads to quite reasonable fits (see Fig.\ \ref{fig6}). 

\begin{figure}[t]
\caption{Convergence with number of particle states of the RPA correction
energy $\Delta_{\rm RPA}$. The extrapolated line as well as the exact 
value (horizontal line) are also shown. The exact values for $N=6$ and 12 
have been taken from the Monte Carlo results of Ref.\ \onlinecite{Pederiva}.}
\label{fig6}
\end{figure}

To assess the quality of $E_{\rm RPA}$ we compare next with available
exact energies, i.e., with {\em ab initio} solutions of the many-body 
Schr\"odinger equation. In fact, the studied cases were chosen in part 
to allow for this comparison. 
For the $N=2$ dot the exact solution can be obtained
by introducing center of mass and relative coordinates. Only the relative 
coordinate problem needs to be solved numerically and, because of the 
circular symmetry, it reduces to an straightforward 1d radial problem.
We have solved this equation using a uniform radial grid and standard
integration methods (see also Refs.\ \onlinecite{YL2,Sim} for recent
discussions of the 2-electron dot exact solution as well as 
Ref.\ \cite{Yan02} for the application of the projection methods to this 
system). 
For the larger dots we shall 
compare with the Monte Carlo results of Pederiva et al.\ \cite{Pederiva}.  

Figure \ref{fig7} presents a systematics for the $N=2$ dots with different values
of the interaction-confinement ratio. As anticipated in Sec.\ IV-A,
$E_{\rm RPA}$ lies below the exact value, a manifestation of an overestimated 
correlation energy in RPA. Nevertheless, for most of the results the RPA  
clearly improves the HF energy, since $E_{\rm RPA}$ is closer to the 
exact value than $E_{\rm HF}$. This is not the case, however, when $R_W=1$; 
the exact result lying almost halfway of RPA and HF in this instance.
For $N=6$ and 12 the results in Fig.\ \ref{fig6} increase towards the exact values. 
The extrapolation in these systems leads to
$\Delta_{\rm RPA}(6)=1.27 E_0$ and 
$\Delta_{\rm RPA}(12)=3.21 E_0$; 
to be compared with the diffusion Monte Carlo values 
$\Delta_{\rm MC}(6)=1.13E_0 $ and 
$\Delta_{\rm MC}(12)=2.64E_0$. The RPA overestimation of correlation 
energy is thus of $\approx 12$\% 
and $\approx 21$\% 
for 6 and 12, respectively. The larger difference for the 12-electron dot
could be partly attributed to the extrapolation procedure being based 
on a relatively small space.  
The deviations with respect to the exact values for $N=2$ are not surprising 
since RPA and HF approaches are many-body theories aiming at an accurate 
description of large enough systems.

\begin{figure}[t]
\caption{Results for the 2-electron dot as a function of $R_W$.
Right scale corresponds to the occupation numbers of the 
HF hole state (solid symbols) 
while left scale indicates the 
ground state energies in HF (circles), RPA (squares) and exact solution 
(triangles). See the discussion about occupation numbers see in Sec.\ IV-C}
\label{fig7}
\end{figure}

\subsection{Occupation numbers and expectation value}

In order to construct the RPA ground state we use Eq.\ (\ref{cond1}) 
and the condition
\beq
\label{lz}
L_z|0\rangle=0\; ,
\eeq
that insures the rotational invariance of the ground state.
One seeks solutions of the form \cite{Rowe,RS,BR86}
\beq
\label{gs0}
|0\rangle=N_0\,\, e^S\,|v\rangle
\eeq
with the $S$ operator involving the creation of two bosons,
\beq
\label{gs1}
S=\frac{1}{2}\sum_{mi\,nj}{
Z_{mi\,nj}\, b_{mi}^{\dagger}b_{nj}^{\dagger}}\; .
\eeq
In Eqs.\ (\ref{gs0}) and (\ref{gs1}) $N_0$ is a normalization constant
and the matrix $Z_{mi\,nj}$ is, 
in general, complex and symmetric in the boson indexes, i.e.,
$Z_{mi\,nj} = Z_{nj\,mi}$.

Using the general indentity
\beq
F e^S = e^S \left( F + [\,F, S\,] + \frac{1}{2} \left[\, [\,F, S\,], S\, \right] + 
\dots \right)
\; 
\eeq
for the operators $F\equiv O_{\lambda}$ and $F\equiv L_z$; as well as 
Eq.\ (\ref{cond1}), Eq.\ (\ref{lz}) and the QBA one finds  
\beq
\label{req}
\left(\, F + [\,F, S\,]\, \right) \, |v\rangle = 0\; .
\eeq
For the case of non-broken symmetry the above requirement leads to the 
following equations for the $Z_{mi\,nj}$ coefficients
\beq
\label{r1}
{Y_{mi}^{(\lambda)*}} = 
\sum_{nj}Z_{mi\,nj}\,{X_{nj}^{(\lambda)*}} \;.
\eeq
When the mean-field breaks rotational symmetry 
we have to complement the above system of equations for vibrons, which are
reduced by one equation, with the additional condition for the spurious mode
\beq
\label{r2}
\ell_{mi}^{(z)} = - \sum_{nj}Z_{mi\,nj}\,{\ell_{nj}^{(z)*}}\; .
\eeq

\begin{figure}[t]
\centerline{\psfig{figure=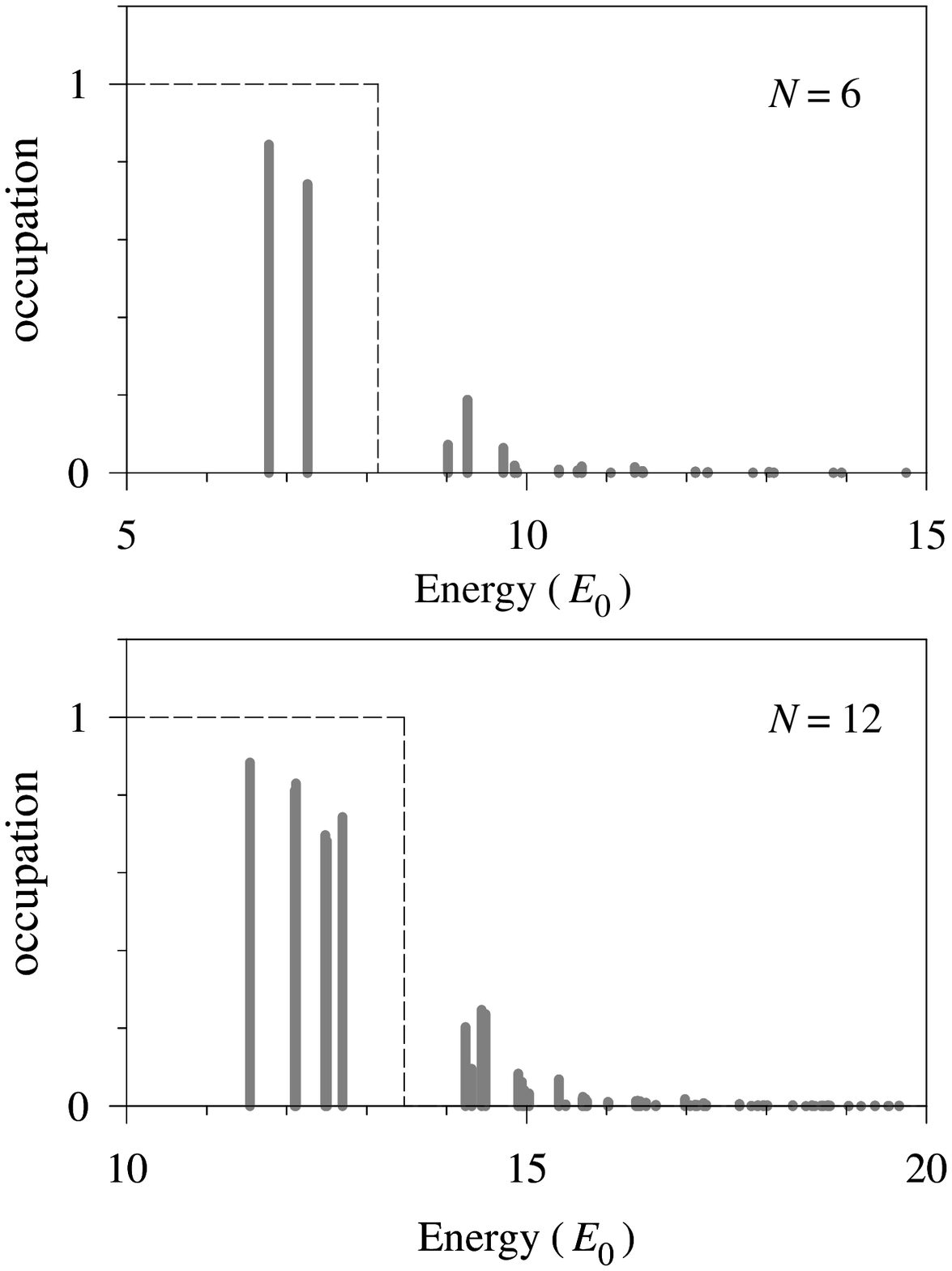,width=3.2in,clip=}}
\caption{ Occupation numbers within RPA of the HF single-particle 
orbitals. The step function shown with a dashed line separates
fully occupied from unoccuppied HF orbitals in the mean field
picture.}
\label{fig8}
\end{figure}

Our interest is focused on the occupation numbers of the HF orbitals
in the correlated ground state,  
$\langle 0| a_m^\dagger a_m |0\rangle$ and   
$\langle 0| a_i^\dagger a_i |0\rangle$. 
Using 
\bea
\mbox{}[a_m^\dagger a_m , b_{nj}^{\dagger}] &=& 
\delta_{mn}b_{mj}^{\dagger}\;\; ,\nonumber\\
\mbox{}[a_i^\dagger a_i , b_{nj}^{\dagger}] &=& 
\delta_{ij}b_{ni}^{\dagger}
\eea
one obtains
\bea
\langle 0 | a_m^\dagger a_m |0 \rangle &=&  
\sum_{inj} Z_{mi\,nj}\,  \langle 0 | b_{mi}^{\dagger} b_{nj}^{\dagger}
|0 \rangle \nonumber\\
\langle 0 | a_i^\dagger a_i |0\rangle &=& 1-
\sum_{mnj} Z_{mi\,nj}\,  \langle 0 | b_{mi}^{\dagger}b_{nj}^{\dagger}|0\rangle\; .
\eea
By substituting in the above expressions the expansion  
of the boson operators  $b_{mi}^{\dagger}$ in terms of 
vibrons and spurious modes, Eq.\ (\ref{def}), we finally obtain
\bea
\label{pp}
\langle 0| a_m^{\dagger}a_m|0\rangle &=& 
\frac{1}{2}\sum_{i\lambda}|Y_{mi}^\lambda|^2
 + \langle 0|\Phi^2 |0\rangle \sum_{i}|\ell_{mi}^z|^2 \qquad\\
\label{hh}
\langle 0 | a_i^\dagger a_i |0\rangle &=& 
1-\frac{1}{2}\sum_{m\lambda}|Y_{mi}^\lambda|^2 \nonumber\\
&-& \langle 0 |\Phi^2 |0\rangle \sum_{m}|\ell_{mi}^z|^2
\eea
These equations generalize the result known in the literature \cite{Rowe} 
by introducing an additional term related to the canonical variables
of the spurious mode $\{L_z, \Phi\}$.
It should be noted that the factor $1/2$ in Eqs.\ (\ref{pp}) and (\ref{hh}) 
is introduced following Ref.\ \onlinecite{14}, where the occupation 
numbers were calculated using Fermionic anticommutator rules, without 
referring to the QBA.

The above discussion can be easily extended to obtain the expectation 
value of any 1-body operator such as, e.g., the particle density
\beq
\hat{\rho}({\bf r})=
\sum_{\alpha\beta}{
\rho_{\alpha \beta}({\bf r})\,
a_{\alpha}^{\dagger}a_\beta}\; ,
\eeq
where indexes $\alpha$ and $\beta $ run over all the HF set of orbitals
and
$\rho_{\alpha \beta}({\bf r})=\varphi^*_\alpha({\bf r})\varphi_\beta({\bf r})$ 
(with $\varphi_\alpha$ the HF wave functions).
Omitting for the sake of presentation the spatial dependence, one can 
write the RPA ground state density as
\beq
\label{dens}
\langle 0 |\hat\rho|0\rangle = \sum_{i}\rho_{ii} - \Delta \rho_{h} 
+ \Delta \rho_{p}\; 
\eeq
with
\bea
\label{ro1}
\Delta \rho_{h} &=& \frac{1}{2} \sum_{\lambda ijm} \rho_{ij}\,
{Y_{mi}^{(\lambda)*} }\, Y_{mj}^{(\lambda)} \nonumber\\
&+& \langle 0 |\Phi^2 |0\rangle \,\sum_{ijm} \rho_{ij}\, 
\ell_{mj}^{(z)*} \ell_{mi}^{(z)} \\
\label{ro2}
\Delta \rho_{p} &=& \frac{1}{2}\sum_{\lambda mnj} \rho_{mn}\,
Y_{mj}^{(\lambda)}\, {Y_{nj}^{(\lambda)*}} \nonumber\\
&+& \langle 0 |\Phi^2 |0\rangle\, \sum_{mnj}\rho_{mn}\,
{\ell_{mj}^{(z)*}}\, \ell_{nj}^{(z)}
\eea

The occupation numbers Eqs.\ (\ref{pp}) and (\ref{hh}), and 
density expectation value Eq.\ (\ref{dens}) depend on the
matrix element $\langle 0 |\Phi^2 |0\rangle$. Involving the two-body 
operator $\Phi^2$ this matrix element is not determined within the 
QBA \cite{MW69}. 
Therefore, we propose the following procedure to fix its value. 
Treating the unknown matrix element as a parameter 
($\gamma\equiv\langle 0 |\Phi^2 |0\rangle$) we 
introduce the function
\beq
F(\gamma)=\int d{{\bf r}}\, 
\left[\, \rho(\gamma,{\bf r})-{\tilde{\rho}}(\gamma, r)\,\right]^2
\eeq
where $\rho(\gamma,{\bf r})$ refers to 
the RPA density Eq.\ (\ref{dens}) and
\beq
{\tilde{\rho}}(\gamma, r) = 
\frac{1}{2\pi}\int_0^{2\pi}d\theta\, \rho(\gamma, {\bf r})
\eeq
is the corrresponding angular average. Since $|0\rangle$ is rotationally
invariant, it is natural to require the physical value $\gamma=\gamma_0$
to be a minimum of $F(\gamma)$. Therefore, it fulfills
the condition
\beq
\left.\frac{dF(\gamma)}{d\gamma}\right\vert_{\gamma=\gamma_0}=0\; .
\eeq
Resolving this equation we obtain the unknown quantity
as
\beq
\gamma_0=-\frac{A}{B}\; , 
\eeq
where
\bea
A&=&\int d{\bf r}\sum_{k=0}^2 \sum_{m=1}^2(-1)^{k+m}
(a_k-{\tilde{a}}_k)(b_m-{\tilde{b}}_m)\; ,\quad\nonumber\\
B&=&\int d{\bf r}\left(\sum_{m=1}^2(-1)^m(b_m-{\tilde{b}}_m)\right)^2\; ,
\eea
and we have defined the $a$ and $b$ coefficients according to
Eqs.\ (\ref{ro1}) and (\ref{ro2}) as
\bea
\Delta\rho_h &\equiv&a_1+\gamma b_1\nonumber\\
\Delta\rho_p &\equiv&a_2+\gamma b_2\nonumber\\
\label{aabb}
a_0&\equiv&\sum_i\rho_{ii}\; .
\eea
The tilde coefficients are defined through analogous equations 
with the circularly averaged densities.

\subsection{Numerical expectation values}

\begin{figure}[t]
\centerline{\psfig{figure=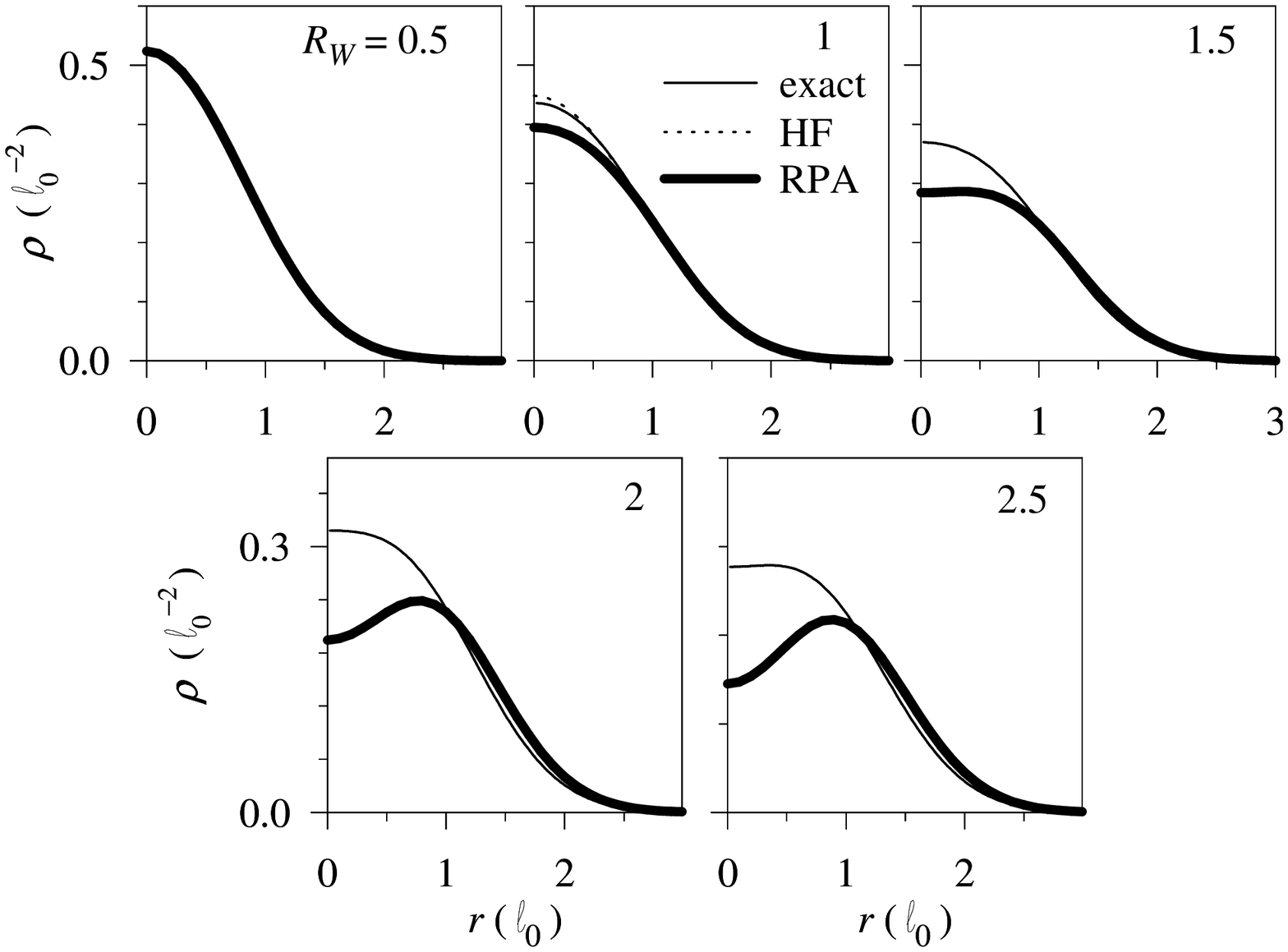,width=3.2in,clip=}}
\caption{Comparison of RPA and exact radial densities for the $N=2$ dot 
with different $R_W$ parameters. In the $R_W=0.5$ and 1 cases the HF
density, which is circular, is also displayed.}
\label{fig9}
\end{figure}

The above discussed Fig.\ \ref{fig7} contains also the numerical values of
the hole state occupation for $N=2$ --there is only one 
occuppied orbital (spin degenerate) in this case. The occupation is close to 1
for weak interaction (low $R_W$) and it decreases in an important way 
as the interaction strength is increased. This proofs that the RPA ground state
deviates from the HF one when interaction becomes strong and correlations 
allow particles and holes to be found above and below the Fermi level,
respectively. The occupations for the other two dots analyzed in this work
are shown in Fig.\ \ref{fig8}.
In these cases we have two (for $N=6$) and four (for $N=12$) different
hole levels with occupations varying with their respective energies. 
In general, the RPA occupation
numbers show a smoother distribution than the HF ones, the maximum 
variations occurring nearby the Fermi level.    

The RPA density is compared with the exact one in Fig.\ \ref{fig9}, 
for the 2-electron dot. For all $R_W$ parameter values the RPA density is 
circularly symmetric; fulfilling the symmetry restoration discussed above. 
Therefore, in  Fig.\ \ref{fig9} we focus exclusively on the radial dependence.  
While the dot edge is well reproduced, a conspicuous feature
is the underestimation of the central density by the RPA,
specially at large $R_W$ values. Our result is in qualitative agreement 
with that of Reinhard \cite{Reinhard} for electrons in jellium spheres
(where a circularly symmetric mean field was imposed). As shown by Reinhard the RPA 
correctly describes the low $q$ components (large $r$'s) of the form factor
$F(q)=\int{dr \exp{(iqr)\rho(r)}}$, but it fails for the large
$q$ contributions (low $r$'s). 
The overcorrection of both central density and correlation energy 
seem thus to be peculiarities of the RPA.

Analogous results for the 6- and 12-electron dots are presented in 
Fig.\ \ref{fig10}. An excellent restoration of the circular symmetry is obtained 
for the $N=6$ dot --we remind the reader that the corresponding HF density was 
shown in Fig.\ \ref{fig2}. For $N=12$ the RPA density, though more circular than the 
HF one, still has some residual deformation. This can be surely attributed
to incompleteness of the RPA space considered in the numerical calculation.
In fact, for this dot the spurious mode separation (Fig.\ \ref{fig4}), the 
ground state energy (Fig.\ \ref{fig6}), and the density (Fig.\ \ref{fig10})
are all indicating that convergence with number of particle states 
is rather slow and, therefore, very difficult to be achieved in a numerical 
calculation. We emphasize that circular symmetry can be restored within
the RPA only if the contribution from the spurious mode is taken into
account.

\begin{figure}[t]
\caption{RPA symmetry restoration of the HF densities displayed
in Fig.\ \ref{fig2}.}
\label{fig10}
\end{figure}

\section{Summary and outlook}

The RPA is a general framework where mean field theory can be 
improved. Focussing on two-dimensional quantum dots we have reviewed the 
description of excited and ground states within the RPA. Some of the results 
known in the literature have been generalized to include broken symmetries at the 
mean field level. Quite importantly, the RPA provides a physical interpretation 
of this broken-symmetry phenomenon: mean-field solutions lacking one 
symmetry of the Hamiltonian represent, according to the RPA,
true {\em internal} deformations of the system having an
associated collective motion at vanishing energy (the spurious RPA mode).
The ground state energy, occupation numbers, as well as general 
expectation values with 1-body operators, including the spurious-mode 
contribution, have been obtained. We proposed the procedure to restore
the rotational symmetry broken at the mean field, which can be extended
for other symmetry breaking cases.

Small quantum dots have provided us a good scenario  
to assess the above properties of the RPA by numerical calculations.
For large enough values of the interaction-confinement ratio $R_W$ the 
HF mean field breaks circular symmetry; the electrons being localized in 
specific geometric distributions. 
In these cases we obtain an spurious 
RPA mode related to the global rotation, with its associated moment of 
inertia. 
The corresponding rotational spectra
may be associated with the rotation of a Wigner molecule \cite{Koon}.
A systematics with $R_W$ for the 2-electron dot has been presented, 
while for 6- and 12-electron dots the calculations have been performed 
with an $R_W$ value suggested by the experiments.

When comparing with exact results, the RPA generally overestimates 
the correlation energy and the central dot densities by an amount
which depends on the electron number and the Hamiltonian parameters.
Another general behaviour of RPA is the slow convergence with 
space dimensions, requiring in practical applications a very high number of
empty HF orbitals. For systems with many electrons this renders the numerical
calculation quite a formidable task. 
In spite of these difficulties the RPA corrections essentially improve the 
HF energies with respect to exact values. 
It should be noted that the QBA only approximately satisfies the Pauli principle
(see Eq.(\ref{qba})). To solve this deficiency of the QBA could be one way
to improve the description of the ground state.
On the other hand, a more systematic treatment of the 
ground state energies would require, probably, alternative many body 
techniques. 

The exact spin symmetry has been imposed on the single-particle states 
in this work, both at the HF
and RPA levels. An interesting extension is, therefore, to
relax this constraint allowing the HF orbitals to be general two-component spinors.
The allowed symmetry breaking would eventually lead to ground states
consisting of vectorial spin textures,
lacking a single quantization axis. This type of states have been predicted 
for quantum dots in large magnetic fields and the question arises
what collective spurious modes will be associated with them. This 
and related issues are left for a future work.

\begin{acknowledgments}
This work was supported by Grant No.\ BFM2002-03241 
from DGI (Spain). R. G. N. gratefully acknowledges support from the 
Ram\'on y Cajal programme (Spain).
\end{acknowledgments}



\begin{thebibliography}{99}
\bibitem{not2} We shall not discuss the ${\bf S}^2$ symmetry restoration
in the present work. Involving a two-body operator, this symmetry is not
automatically included in the RPA. It can be dealt, however, with projection 
methods \cite{Yan02}.
\end{thebibliography}
\end{document}